\documentclass[aps,pra,twocolumn,10pt,showpacs,preprintnumbers,amsmath,amssymb,floatfix,longbibliography]{revtex4-2}

\usepackage{graphicx}
\usepackage{dcolumn}
\usepackage{bm}
\usepackage{color}
\usepackage{mathrsfs}

\newcommand{\bra}[1]{\langle #1|}
\newcommand{\ket}[1]{|#1\rangle}

\newcommand{\affA}{\vspace{1ex}$^1$Institute for Theoretical Physics, Institute of Physics, University of Amsterdam,\\ Science Park 904, 1098 XH Amsterdam, the Netherlands}
\newcommand{\affB}{$^2$QuSoft, Science Park 123, 1098 XG Amsterdam, the Netherlands}
\newcommand{\affC}{$^3$Zentrum f\"ur Optische Quantentechnologien, Universit\"at Hamburg, Luruper Chaussee 149, 22761 Hamburg, Germany}
\newcommand{\affD}{$^4$Van der Waals-Zeeman Institute, Institute of Physics, University of Amsterdam, 1098 XH Amsterdam, the Netherlands}

\usepackage[normalem]{ulem}
\usepackage{physics}
\usepackage{multirow}

\usepackage{makecell}

\usepackage{hyperref}
\usepackage{xcolor}
\hypersetup{
    colorlinks,
    linkcolor={red!70!black},
    citecolor={green!65!black},
    urlcolor={blue!80!black}
}

\begin{document}

\title{Phonon-mediated quantum gates in trapped ions coupled to an ultracold atomic gas}

\author{Lorenzo Oghittu$^{1,2}$, Arghavan Safavi-Naini$^{1,2}$, Antonio Negretti$^{3}$ and Rene Gerritsma$^{1,4}$}
\affiliation{\affA,\affB,\affC,\affD}
\date{\today}

\begin{abstract}
We study the dynamics of phonon-mediated qubit-qubit interactions between trapped ions in the presence of an ultracold atomic gas. By deriving and solving a master equation to describe the combined system, we show that the presence of the atoms causes the quantum gate quality to reduce because of motional decoherence. On the other hand, we calculate that the gas may be used to keep the ion crystal cold in the presence of external heating due to electric field noise. We show that tuning the atom-ion scattering length allows one to tune the cooling rate of the ions and would make it possible to temporarily reduce the effects of the gas during a quantum gate while keeping the ions cold over long timescales. In this way, the trapped ion quantum computer may be buffer gas cooled. The system may also be used for quantum-enhanced measurements of the atom-ion interactions or properties of the atomic bath. 
\end{abstract}

\maketitle

\section{Introduction}
Over the last decade, rapid progress has occurred in controlling laboratory systems that explore the quantum dynamics of mixtures of ultracold atoms and trapped ions~\cite{Cot'e:2002,Idziaszek:2007,Grier:2009,Zipkes:2010,Schmid:2010,Tomza:2017cold,Meir:2016,Haze:2018,Feldker:2020,Lous2022}. These systems offer new opportunities in studying quantum impurity physics, ultracold collisions and quantum chemistry~\cite{Ratschbacher:2012,Harter:2012,Haze:2015_,Joger:2017,Sikorsky:2018,Dieterle_PRL21,Katz:2022}. On the theoretical side, efforts have been made to study and characterize ionic polarons~\cite{Schurer_PRL17,Christensen_PRL21,Astrakharchik_CP21} and bath-mediated interactions between charged impurities~\cite{Ding_PRL22,Astrakharchik_NC23,Olivas_PRA24}. Moreover, systems involving ionic impurities in ultracold gases have been proposed as quantum simulators and sensors~\cite{Bissbort_PRL13,Negretti_PRB14,Jachymski_PRR20,Oghittu_PRR22}. Combining atom-ion systems with precise quantum control of individual trapped ions opens the possibility to study the decoherence of non-classical states of ion qubits and motion~\cite{Leibfried:1996,Monroe:1996} as well as phonon-mediated interactions between ion qubit states~\cite{Molmer:1999,Leibfried:2003a,Benhelm:2008b} inside a bath of ultracold atoms. These mixed atom-ion systems may find applications in quantum-enhanced sensing of atom-ion interactions~\cite{Katz:2022} and the properties of the atomic bath~\cite{Hirzler:2022}. Furthermore, the ultracold atomic gas may be used to cool the trapped ions, with recent results showing that buffer gas cooling even outperforms laser (Doppler) cooling~\cite{Feldker:2020,Schmidt:2020}. Finally, the observation of atom-ion Feshbach resonances has greatly enhanced the possibilities of atom-ion quantum systems~\cite{Weckesser2021oof}. In particular, these resonances allow for controlling the atom-ion interactions by simple tuning of an external magnetic field, in analogy to the case of Feshbach resonances between neutral atoms~\cite{Chin:2010}.

In this paper, we consider the case where ionic qubits are coupled via laser-driven phonon-mediated interactions and are coupled to a quantum bath of ultracold atoms. We study how the atoms change the dynamics of this system and affect the quality of quantum gates between the qubits. We find that the main source of error occurs due to decoherence of the ion motion in phase space during the gate. We consider the benefits of the atomic bath to keep the ions cool in the presence of an external heating source due to electric field noise. We study the role of the atom-ion s-wave scattering length, which may be tuned using Feshbach resonances. We find that it is possible to switch the interactions in such a way that either the adverse effects of the bath on the quantum gate or the negative effects of heating may be mitigated. These results show that it is possible to buffer gas cool the trapped ion quantum system~\cite{Daley:2004}, while maintaining the quality of quantum operations. Moreover, our model involves a setup that is experimentally accessible to several groups working on hybrid atom-ion systems, and therefore paves the way for the study of combined systems of ultracold atoms and quantum gates.

\begin{figure}[h!]
    \centering
    \includegraphics[width=.5\textwidth]{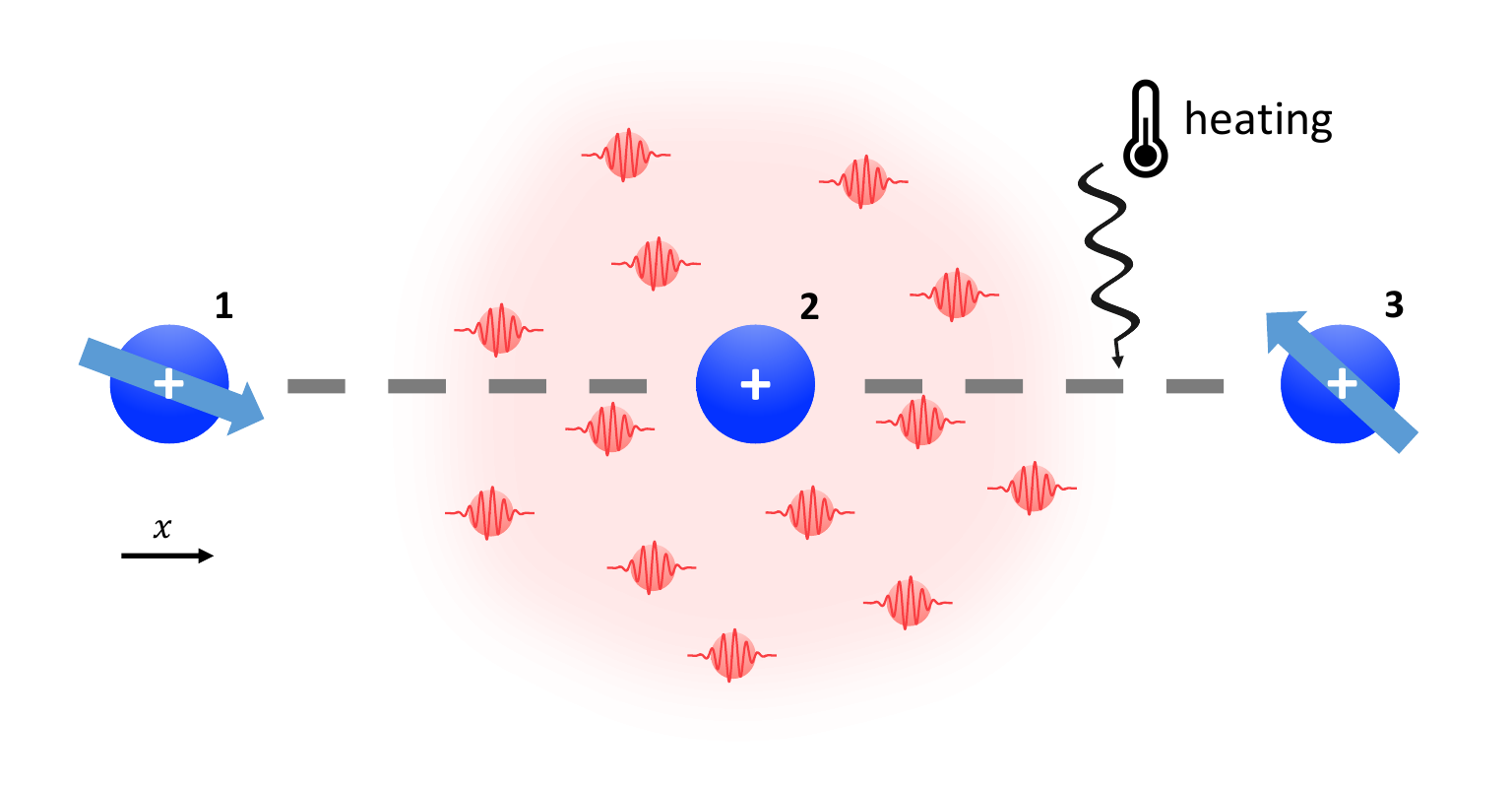}
    \caption{Scheme of the system. The three ions are described as a one dimensional chain whose motion is decomposed into three normal modes of oscillation (dashed lines). Ions 1 and 3 are considered as two-level spin systems. Ion 2 is immersed in a bath of ultracold atoms (red particles) which are assumed to be effectively free}. A thermal bath at room temperature (wavy line) is coupled to the center-of-mass motion of the ion chain.
    \label{fig:scheme}
\end{figure}

The paper is organized as follows: in Sec.~\ref{sec:system} we describe the Hamiltonian of the system, which is used in Sec.~\ref{sec:master_equation} to derive a master equation for the ion chain. In Sec.~\ref{sec:results} we discuss the results obtained by numerically solving the master equation. Finally, conclusions and outlook are provided in Sec.~\ref{sec:conclusions}.

\section{System description}
\label{sec:system}
We consider the scheme depicted in Fig.~\ref{fig:scheme}: it consists of a linear crystal of three ions, where the outer two ions experience a spin-dependent force that couples their (internal) spin states to the (external) vibrational phononic modes of the chain~\cite{Molmer:1999}. The entire chain is coupled to an external bath by immersing the central ion in a cloud of ultracold atoms.  Note that the distance between the equilibrium position of the ions for the frequencies considered in this paper (see Tab.~\ref{tab:frequencies}) is approximately $5\,\mathrm{\mu m}$, which is comparable to the size of typical ultracold atomic clouds. Hence, the outer ions can be realistically separated from the cloud by several micrometers, making our model sensible from an experimental point of view. By keeping the two qubits out of the gas, we avoid possible spin changing collisions between atoms and ions such as spin relaxation. The latter, in which the total spin is not conserved, has been observed in atom-ion experiments~\cite{Ratschbacher:2013,Tscherbul:2016,Sikorsky:2018,Fuerst:2018:spin}. Additionally, the center-of-mass (c.m.) mode is subjected to external heating due to electric field noise at room temperature.

In this section we describe the Hamiltonian of the system. In Sec.~\ref{subsec:ion_chain} we focus on the the ion chain and the spin-dependent force acting on the two outer ions, while in  Sec.~\ref{subsec:gas_coupling} we describe the interaction of the central ion with the ultracold quantum gas.

\subsection{Ion chain and spin-dependent force}
\label{subsec:ion_chain}
We consider three identical ions in a linear Paul trap, and we assume the confinement along $y$ and $z$ to be much stronger than the confinement along the direction of the trap axes $x$. In such a configuration, the motion along $y$ and $z$ can be neglected, and the interplay between the harmonic confinement and the Coulomb repulsion along $x$ allows one to describe the three ions as a one-dimensional crystalline chain. The motion of the ions is therefore approximated as a displacement around their equilibrium position and decomposed into three normal modes of oscillation \cite{Raizen_PRA92,James_APB98}. On top of the motional (phononic) degrees of freedom, the two outer ions are equipped with two internal spin states and will therefore be referred to as qubits.
The phononic modes are coupled to the spin states by a spin-dependent force along the direction perpendicular to the chain. This is done with lasers in a Raman scheme with wave vector $k_R$~\cite{Leibfried:2003a,Sorensen_PRL99,Sorensen_PRA00}. When the displacement of the ions is much smaller than the wavelength of the lasers, the corresponding Hamiltonian can be expanded to lowest order in the Lamb-Dicke parameters $\eta_\mu=k_R\sqrt{\hbar/(2M\omega_\mu)}$, with $\omega_\mu$ the frequency of mode $\mu$. In the frame rotating with the Raman beatnote $\omega_R$, we obtain the following spin-phonon Hamiltonian:
\begin{equation}
    \hat{H}_\mathrm{sp}=-\frac{\hbar}{2}\sum_{j=1,3}\sum_{\mu}\Omega_\mu b_{j,\mu}\big(\hat{a}_\mu+\hat{a}_\mu^\dag\big)\hat{\sigma}_j^z,
\label{eq:H_spin-phonon}
\end{equation}
where we performed the rotating wave approximation. Here, $\hat{\sigma}_j^z$ is the $z$ Pauli matrix acting on spin $j$, while $\hat{a}_\mu^{(\dag)}$ annihilates (creates) a phonon of mode $\mu$, whose amplitude vector is defined by $\mathbf{b}_\mu$. The amplitude vectors for the case of three particles in one dimension are reported in Tab.~\ref{tab:modes}. Note that the sum over the ions in Eq.~\eqref{eq:H_spin-phonon} does not include the central ion $j=2$, as this is not affected by the spin-dependent force. Finally, the driving strength of mode $\mu$ is quantified by $\Omega_\mu=F\eta_\mu/k_R$, where $F$ is the magnitude of the spin-dependent force.

In the same frame, the total Hamiltonian of the ion chain with spin-phonon coupling is given by $\hat{H}_\mathrm{chain}=\hat{H}_\mathrm{s}+\hat{H}_\mathrm{p}+\hat{H}_\mathrm{sp}$. The spin Hamiltonian is $\hat{H}_\mathrm{s}=0$, while the phonon Hamiltonian reads
\begin{equation}
    \hat{H}_\mathrm{p}=-\hbar\sum_{\mu}\delta_\mu\hat{n}_\mu
    \label{eq:H_phonon}
\end{equation}
where $\hat{n}_\mu=\hat{a}_\mu^\dag\hat{a}_\mu$ is the number operator corresponding to mode $\mu$ and $\delta_\mu=\omega_R-\omega_\mu$ is the detuning of the Raman beat note from the mode frequency.
\begin{table}
    \centering
    \setlength{\extrarowheight}{1ex}
    \setlength{\tabcolsep}{2ex}
    \begin{tabular}{l l l}
    \hline\hline
        Normal mode & Eigenvector & Eigenvalue \\[1ex]
    \hline
        c.m.: & $\mathbf{b}_\mathrm{c.m.}=(1,1,1)/\sqrt{3}$, & $\lambda_\mathrm{c.m.}=1$ \\[1ex]
        stretching: & $\mathbf{b}_\mathrm{st}=(-1,0,1)/\sqrt{2}$, & $\lambda_\mathrm{st}=3$ \\[1ex]
        wobbling: & $\mathbf{b}_\mathrm{wb}=(1,-2,1)/\sqrt{6}$, & $\lambda_\mathrm{wb}=29/5$\\[1ex]
        \hline\hline
    \end{tabular}
    \caption{Normal modes eigenvectors and eigenvalues for a 1D chain of three ions with equal mass \cite{James_APB98}.}
    \label{tab:modes}
\end{table}
In this paper, we consider the parameters reported in Tab.~\ref{tab:frequencies}.

The time evolution corresponding to the interaction Hamiltonian in Eq.~\eqref{eq:H_spin-phonon} can be computed analytically. This is done by means of the Magnus expansion~\cite{Magnus_CPAM54}, which truncates exactly at second order in this specific case. We have
\begin{equation}
    \begin{split}
        \hat{U}_I(t,0)=&\mathrm{exp}\bigg\{-\frac{i}{\hbar}\int_0^tdt'\,\Tilde{H}_\mathrm{sp}(t')\\
        &-\frac{i}{2\hbar^2}\int_0^tdt'\int_0^{t'}dt{''}\comm{\Tilde{H}_\mathrm{sp}(t')}{\Tilde{H}_\mathrm{sp}(t'')}\bigg\}\\[2ex]
        =&\hat{U}_\mathrm{sp}(t)\hat{U}_\mathrm{ss}(t)
    \end{split}
\label{eq:magnus}
\end{equation}
where $\Tilde{H}_\mathrm{sp}(t)$ is the spin-phonon Hamiltonian in the interaction picture with respect to $\hat{H}_\mathrm{p}$~\footnote{Note that we are combining the transformation to the frame rotating with $\omega_R$ with the one to the interaction picture with respect to $\hat{H}_\mathrm{p}$. This is equivalent to transforming to the interaction picture in the laboratory frame, provided that all Hamiltonian operators are expressed in the same frame.}. The last line shows that the total propagator factorizes as the product of the spin-phonon and spin-spin coupling contributions, which are given, respectively, by
\begin{equation}
    \hat{U}_\mathrm{sp}(t)=\mathrm{exp}\bigg\{\sum_{j=1,3}\sum_\mu\Big(\phi_{j,\mu}(t)\hat{a}_\mu^\dag-\phi_{j,\mu}^*(t)\hat{a}_\mu\Big)\hat{\sigma}_j^z\bigg\},
\label{eq:U_sp}
\end{equation}
and
\begin{equation}
    \hat{U}_\mathrm{ss}(t)=\mathrm{exp}\bigg\{-i\sum_{i,j}J_{i,j}(t)\hat{\sigma}_i^z\hat{\sigma}_j^z\bigg\}.
\label{eq:U_ss}
\end{equation}
Here, we used the definitions
\begin{equation}
    \begin{split}
        \phi_{j,\mu}(t)&=\frac{\Omega_\mu}{2\delta_\mu} b_{j,\mu}\Big(1-e^{-i\delta_\mu t}\Big),\\[2ex]
        J_{i,j}(t)&=\frac{1}{4}\sum_\mu\Omega_\mu^2\frac{b_{j,\mu}b_{i,\mu}}{\delta_\mu^2}\Big(\delta_\mu t-\mathrm{sin}(\delta_\mu t)\Big).
    \end{split}
\label{eq:phi_and_J}
\end{equation}
The term in Eq.~\eqref{eq:U_sp} corresponds to a spin-dependent displacement and it is responsible for the entanglement between spin and motion. When $|\delta_\mu|\gg\Omega_\mu$ for all modes, however, we see from Eq.~\eqref{eq:phi_and_J} that $\phi_{j,\mu}\approx0$ and the evolution reduces to the propagator in Eq.~\eqref{eq:U_ss}. In this regime, the interaction Hamiltonian is effectively described by the spin-spin interaction in the exponent of $\hat{U}_\mathrm{ss}$ at all times. When only the mode $\nu$ is driven substantially, i.e. $\delta_\nu\gtrsim\Omega_\nu$, the other modes can be neglected and $\phi_{j,\nu}=0$ when $t$ is an integer multiple of $2\pi/\delta_\nu$. At these values of time, the motion is disentangled from the spin, meaning that the internal state does not depend on the external vibrational state \cite{Sorensen_PRA00,Kim:2009}. Note however, that for larger vibrational state occupation the Lamb-Dicke approximation breaks down. This  causes a reduction in the quantum gate quality. In practice, high fidelity quantum gate operation relies on sub-Doppler cooling~\cite{Kirchmair:2009,Ballance:2016,Gaebler:2016} and minimal background heating~\cite{Brownnutt_RMP15}.

\subsection{Coupling to the quantum gas}
\label{subsec:gas_coupling}
\textit{Atom-ion potential –} The atom-ion interaction is represented asymptotically at large separation by the polarization potential
\begin{equation}
    V_\mathrm{ai}(r)=-\frac{C_4}{r^4}, \qquad C_4=\frac{\alpha e^2}{2}\frac{1}{4\pi\epsilon_0},
\end{equation}
where $\alpha$ is the static polarizability of the atom, $e$ is the electronic charge and $\epsilon_0$ is the vacuum permittivity.
The characteristic length and energy scales of the polarization potential are $R^\star=(2\mu C_4/\hbar^2)^{1/2}$ and $E^\star=\hbar^2/[2\mu(R^\star)^2]$, respectively.

For analytical purposes, we consider a regularization of the atom-ion polarization potential \cite{Krych_PRA15}:
\begin{equation}
    V_\mathrm{ai}^r(\mathbf{r})=-C_4\frac{r^2-c^2}{r^2+c^2}\frac{1}{(b^2+r^2)^2}
    \label{eq:V_reg}
\end{equation}
where $b$ and $c$ are parameters that can be associated to the $s$-wave atom-ion scattering length $a_\mathrm{ai}$ and the number of two-body bound states supported by the potential. One useful property of Eq.~\eqref{eq:V_reg} is that its Fourier transform can be computed analytically, allowing the scattering amplitude in the first-order Born approximation to be written as follows:
\begin{equation}
    f(q)=\frac{c^2\pi(R^\star)^2}{(b^2-c^2)^2q}\bigg\{e^{-bq}\bigg[1+\frac{(b^4-c^4)q}{4bc^2}\bigg]-e^{-cq}\bigg\}.
    \label{eq:scatt_amplitude}
\end{equation}
This will be used in the following sections.\\

\textit{Coupling Hamiltonian –} The central ion (labeled with $j=2$) is immersed in a homogeneous ultracold gas of bosons or fermions confined in a box of length $L$, which is assumed to be much larger than any of the other lengths involved in the description of the ionic impurity in the ultracold bath. This is justified by the low frequencies of typical atomic traps compared to ion traps in experiments. The bath Hamiltonian is given by
\begin{equation}
    \hat{H}_\mathrm{bath}=\int_{\mathbb{R}^3}d\mathbf{r}_b\hat{\Psi}^\dag_b(\mathbf{r}_b)\bigg[\frac{\hat{\mathbf{p}}_b^2}{2m}+\frac{g}{2}\hat{\Psi}^\dag_b(\mathbf{r}_b)\hat{\Psi}_b(\mathbf{r}_b)\bigg]\hat{\Psi}_b(\mathbf{r}_b)
\end{equation}
where the subscript $b$ indicates that the quantities refer to the bath, $\hat{\Psi}^\dag_b$ and $\hat{\Psi}_b$ are the bath field operators and $g$ is the atom-atom interaction strength.

In the laboratory frame, the Hamiltonian accounting for the coupling between the ion chain and the atomic bath reads
\begin{equation}
    \hat{H}_\mathrm{int}=\int_{\mathbb{R}^3}d\mathbf{r}_b\hat{\Psi}^\dag_b(\mathbf{r}_b)V_\mathrm{ai}(\mathbf{r}_b-\hat{\mathbf{r}}_2)\hat{\Psi}_b(\mathbf{r}_b),
    \label{eq:H_int}
\end{equation}
where $\hat{\mathbf{r}}$ is the position operator of the ion. Following the procedure and notation adopted in Ref.~\cite{Oghittu_PRA21}, we express the fluctuations around the condensate in terms of Bogoliubov modes. Let us note that we will refer to phonons in the bath as atomic phonons, in order to avoid confusion with the phonons of the vibrational modes of the ion chain. In the bosonic case, we obtain the following Hamiltonian
\begin{equation}
    H_\mathrm{int}^\mathrm{B}=\hbar\sum_\mathbf{q}\big(\hat{S}_\mathbf{q}\hat{b}_\mathbf{q}+\hat{S}_\mathbf{q}^\dag\hat{b}_\mathbf{q}^\dag\big)+\hbar\sum_{\mathbf{q},\mathbf{q'}}\hat{S}_{\mathbf{q},\mathbf{q'}}\hat{b}_\mathbf{q}^\dag\hat{b}_\mathbf{q'}
    \label{eq:H_int_B}
\end{equation}
where $\hat{b}_\mathbf{q}^{\dag}$ and $\hat{b}_\mathbf{q}$ are the creation and annihilation operators of an atomic phonon excitation with momentum $\mathbf{q}$ in the bosonic bath and obey the commutation relation $[\hat{b}_\mathbf{q},\hat{b}_\mathbf{q'}^\dag]=\delta_{\mathbf{q},\mathbf{q'}}$. Moreover, we defined 
\begin{equation}
    \hat{S}_\mathbf{q}=\frac{\sqrt{n_0L^3}}{\hbar}e^{i\mathbf{q}\cdot\hat{\mathbf{r}}}c_\mathbf{q},\qquad \hat{S}_{\mathbf{q},\mathbf{q'}}=\frac{1}{\hbar}e^{i(\mathbf{q'}-\mathbf{q})\cdot\hat{\mathbf{r}}}c_{\mathbf{q'}-\mathbf{q}},
    \label{eq:S_operators}
\end{equation}
where $n_0$ is the density of the condensate, and the coefficient $c_\mathbf{q}$ is related to the scattering amplitude $f(q)$ of the atom-ion potential by
\begin{equation}
    c_\mathbf{q}=-\frac{2\pi\hbar^2}{\mu L^3}f(q).
    \label{eq:c_coefficient}
\end{equation}
In the case of the regularized polarization potential that we consider in this paper, $f(q)$ is given in Eq.~\eqref{eq:scatt_amplitude}. 

We note that the definitions in Eq.~\eqref{eq:S_operators} and the second sum in Eq.~\eqref{eq:H_int_B} are obtained by considering that only atomic phonons with energy comparable to $\hbar\omega_\mu$ will couple to the ion motion. For instance, let us consider the center-of-mass frequency $\omega_\mathrm{c.m.}=2\pi\cdot500\,\mathrm{kHz}$, corresponding to an atom velocity $\sqrt{2\hbar\omega_\mathrm{c.m.}/m}\simeq0.24\,\mathrm{m/s}$. This is much faster than the speed of sound in the $^7$Li condensate, which for a condensate density $n_0=10^{14}\,\mathrm{c.m.}^{-3}$ is on the order of $0.005\,\mathrm{m/s}$. Therefore, we can safely assume a particle-like dispersion relation $\hbar\omega_\mathbf{q}=\hbar^2q^2/(2m)$ for the atomic phonons, and treat the bath as a non-interacting Bose gas. We refer to Ref.~\cite{Oghittu_PRA21} for the general expressions before the particle-like approximation.

In the case of a Fermi gas, the coupling Hamiltonian simply becomes
\begin{equation}
    H_\mathrm{int}^\mathrm{F}=\hbar\sum_{\mathbf{q},\mathbf{q'}}\hat{S}_{\mathbf{q},\mathbf{q'}}\hat{f}_\mathbf{q}^\dag\hat{f}_\mathbf{q'}
    \label{eq:H_int_F}
\end{equation}
where the operators $\hat{f}_\mathbf{q}^\dag$ and $\hat{f}_\mathbf{q}$ create and annihilate a fermion with momentum $\mathbf{q}$ and obey the anticommutation relation $\{\hat{f}_\mathbf{q},\hat{f}_\mathbf{q'}^\dag\}=\delta_{\mathbf{q},\mathbf{q'}}$.

The derivation of the master equation (see Sec.~\ref{sec:master_equation}), requires the system-bath Hamiltonian to be expressed in the same rotating frame of the system Hamiltonian. This is obtained by transforming Eq.~\eqref{eq:H_int_B} and Eq.~\eqref{eq:H_int_F} with the operator $\hat{\mathcal{U}}_R(t)=\mathrm{exp}[-i\omega_R\sum_\mu\hat{n}_\mu t]$. Let us consider the coupling to the condensed fraction of a Bose gas, which is given by the first sum in Eq.~\eqref{eq:H_int_B}. The transformation only acts on the phonon operators, and the Hamiltonian has the same form, but with the time-dependent operator
\begin{equation}
    \hat{S}_\mathbf{q}(t)=\frac{\sqrt{n_0 L^3}}{\hbar}e^{iq_x\hat{x}_R(t)},
    \label{eq:S_R}
\end{equation}
where we made use of the fact that the chain oscillates along the $x$ direction and we defined the position operator of the central ion in the rotating frame as
\begin{equation}
    \hat{x}_R(t)=\sum_{\mu}b_{2,\mu}\lambda_\mu\Big(\hat{a}_\mu e^{-i\omega_Rt}+\hat{a}_\mu^\dag e^{i\omega_Rt}\Big),
\end{equation}
with $\lambda_\mu=\sqrt{\hbar/(2M\omega_\mu)}$ the length associated to mode $\mu$. The coupling to the non condensed Bose gas, represented by the double sum in Eq.~\eqref{eq:H_int_B}, and to the Fermi gas, reported in Eq.~\eqref{eq:H_int_F}, is transformed to the rotating frame in an analogous manner.

We finally remark that working in the frame rotating with $\omega_R$ allows the ion chain Hamiltonian to be time-independent, as long as the rotating-wave approximation holds. This simplifies the derivation of the master equation, although a time dependence is gained by the coupling to the bath. We refer to Sec.~\ref{sec:master_equation} for more details.

\section{Master equation}
\label{sec:master_equation}
We describe the evolution of the system with a master equation  \cite{Carmichael}, in analogy to what has been done in Ref.~\cite{Krych_PRA15,Oghittu_PRA21,Oghittu_PRR24}. Here, we only report the most important steps of the derivation, focusing on the differences with respect to the previous works. In Sec.~\ref{subsec:bath_ME} we derive the contribution from the coupling to the quantum gas, while in Sec.~\ref{subsec:heating_ME} the external heating term is considered.

\subsection{Master equation for the ion chain coupled to the ultracold gas}
\label{subsec:bath_ME}
Our open system is represented by the ion chain with spin-phonon coupling, whose Hamiltonian $\hat{H}_\mathrm{chain}$ is defined in Sec.~\ref{subsec:ion_chain} [see Eq.~\eqref{eq:H_spin-phonon} and \eqref{eq:H_phonon}]. As a starting point for the derivation, we consider the Redfield equation:
\begin{equation}
    \frac{d}{dt}\Tilde{\rho}_R(t)=-\int_0^t\frac{dt'}{\hbar^2}\,\mathrm{Tr}_b\Big\{\Big[\Tilde{H}_\mathrm{int}(t),\big[\Tilde{H}_\mathrm{int}(t'),\Tilde{\rho}_R(t)\otimes\hat{B}_0\big]\Big]\Big\}
    \label{eq:Redfield}
\end{equation}
where $\Tilde{\rho}_R$ and $\Tilde{H}_\mathrm{int}$ are, respectively, the reduced density matrix of the system and the system-bath interaction Hamiltonian in the rotating frame and in the interaction picture with respect to $\hat{H}_\mathrm{chain}+\hat{H}_\mathrm{bath}$. Moreover, $\hat{B}_0$ is the density matrix of the bath and $\mathrm{Tr}_b\{\dots\}$ represents the trace over the bath degrees of freedom. We note that Eq.~\eqref{eq:Redfield} relies on the Markov and Born approximations.
Let us now focus on the case where the bath is a Bose-Einstein condensate. By explicitly writing $\Tilde{H}_\mathrm{int}$ and performing the trace, we get
\begin{widetext}
    \begin{equation}
    \begin{split}
        \frac{d}{dt}\Tilde{\rho}_R(t)=-\int_0^tdt'\,\sum_\mathbf{q}\bigg\{\big(n_\mathbf{q}+1\big)\Big(e^{-i\omega_\mathbf{q}(t-t')}&\comm{\Tilde{S}_\mathbf{q}(t)}{\Tilde{S}_\mathbf{q}^\dag(t')\Tilde{\rho}_R(t)}+\mathrm{H.\,c.}\Big)\\
        +&n_\mathbf{q}\Big(e^{-i\omega_\mathbf{q}(t-t')}\comm{\Tilde{\rho}_R(t)\Tilde{S}_\mathbf{q}^\dag(t')}{\Tilde{S}_\mathbf{q}(t)}+\mathrm{H.\,c.}\Big)\bigg\}
    \end{split}
    \label{eq:ME_intpicture}
\end{equation}
\end{widetext}
where $\hbar\omega_\mathbf{q}$ is the energy of the atomic phonons in the condensate, H.\,c. indicates the Hermitian conjugate, $\Tilde{S}_\mathbf{q}$ represents the operator in Eq.~\eqref{eq:S_R} when transformed to the interaction picture and $n_\mathbf{q}=[e^{\beta(\hbar\omega_\mathbf{q}-\mu_\mathbf{B})}-1]^{-1}$ is the Bose-Einstein occupation number obtained from the thermal averages over the state of the bath $\hat{B}_0$ 
\begin{equation}
    \mathrm{Tr}_b\Big\{\hat{b}_\mathbf{q}^\dag\hat{b}_\mathbf{q'}\hat{B}_0 \Big\}=n_\mathbf{q}\delta_{\mathbf{q},\mathbf{q'}},
\end{equation}
with $\mu_\mathrm{B}$ the chemical potential of the gas at temperature $T$ and $\beta=1/(k_BT)$ ($k_B$ the Boltzmann constant).

The next step consists of transforming the master equation back to the Schr\"{o}dinger picture and perform the Lamb-Dicke approximation for small values of $q_x\hat{x}$. As an example, we consider the first commutator in Eq.~\eqref{eq:ME_intpicture}, from which we get
\begin{align}
    \comm{e^{iq_x\hat{x}_R(t)}}{e^{iq_x\hat{x}_R(t',-\tau)}\hat{\rho}_R(t)}&\nonumber \\
        \simeq\comm{\hat{x}_R(t)}{\hat{x}_R(t',-\tau)\hat{\rho}_R(t)}&-\frac{1}{2}\comm{(\hat{x}_R(t))^2}{\hat{\rho}_R(t)}
\end{align}
where we neglected the term proportional to $q_x$ as this vanishes after the sum over $\mathbf{q}$ due to spherical symmetry of the bath. Here, we defined $\tau=t-t'$ and
\begin{equation}
    \hat{x}_R(t,-\tau)=e^{-i\hat{H}_\mathrm{chain}\tau}\,\hat{x}_R(t)\,e^{i\hat{H}_\mathrm{chain}\tau}.
    \label{eq:x_t-tau}
\end{equation}
Because of the time-independence of $\hat{H}_\mathrm{chain}$ (see Sec.~\ref{sec:system}), it is straightforward to calculate the product in Eq.~\eqref{eq:x_t-tau} by means of the Baker-Campbell-Hausdorff identity~\footnote{Baker-Campbell-Hausdorff identity: $e^{i\hat{G}\lambda}\hat{A}e^{-i\hat{G}\lambda}=\hat{A}+i\lambda[\hat{G},\hat{A}]+(i\lambda)^2[\hat{G},[\hat{G},\hat{A}]]/2!+\dots$}. The explicit form of $\hat{x}_R(t,-\tau)$ is reported in Eq.~\eqref{eq:x_t-tau_long}.

We refer the interested reader to Ref.~\cite{Krych_PRA15,Oghittu_PRA21,Oghittu_PRR24} for more details and for the case of a Fermi gas.
We now report the master equation for the ion chain coupled to a Bose-Einstein condensate in the frame rotating with $\omega_R$: 
\begin{widetext}
\begin{equation}
    \begin{split}
        \frac{d}{dt}\hat{\rho}_R=-&\frac{i}{\hbar}\comm{\hat{H}_\mathrm{chain}}{\hat{\rho}_R}\\
        -\Gamma\sum_{\mu}\bigg\{&\Big(n_{q_\mu}\comm{\hat{x}_R}{\big(\hat{\alpha}_\mu+\hat{\alpha}^\dag_\mu\big)\hat{\rho}_R-\hat{\rho}_R\big(\hat{\alpha}_\mu+\hat{\alpha}^\dag_\mu\big)}+\comm{\hat{x}_R}{\hat{\alpha}_\mu\hat{\rho}_R-\hat{\rho}_R\hat{\alpha}_\mu^\dag}\Big)\\
        +\frac{\Omega_\mu}{2\delta_\mu}&\sum_{j=1,3}b_{j,\mu}\Big(\mathrm{cos}\big(\omega_R t\big)\comm{\hat{x}_R}{\hat{\sigma}_j^z\hat{\rho}_R-\hat{\rho}_R\hat{\sigma}_j^z}2h_\mu^{(n_q)}+e^{-i\omega_Rt}\comm{\hat{x}_R}{\hat{\sigma}_j^z\hat{\rho}_R}h_\mu-e^{i\omega_Rt}\comm{\hat{x}_R}{\hat{\rho}_R\hat{\sigma}_j^z}h_\mu\Big)\bigg\},
    \end{split}
    \label{eq:equation_BEC}
\end{equation}
\end{widetext}
where we omitted time dependencies for clarity and we defined $\Gamma=2\pi m \hbar n_0/(3\mu^2)$,
\begin{equation}
    \begin{split}
        &\hat{\alpha}_\mu(t)=b_{2,\mu}\lambda_\mu|f(q_\mu)|^2q_\mu^3\hat{a}_\mu e^{-i\omega_Rt}\\
        &h_\mu=b_{2,\mu}\lambda_\mu\Big(|f(q_\mu)|^2q_\mu^3-|f(q_R)|^2q_R^3\Big)\\
        &h_\mu^{(n_q)}=b_{2,\mu}\lambda_\mu\Big(n_{q_\mu}|f(q_\mu)|^2q_\mu^3-n_{q_R} |f(q_R)|^2q_R^3\Big)
    \end{split}
\end{equation}
and $q_{\mu(R)}=\sqrt{2m\omega_{\mu(R)}/\hbar}$. 

The first line of Eq.~\eqref{eq:equation_BEC} is the unitary evolution of the system in absence of the bath, while the second and third lines describe the dissipative contribution due to the coupling to the Bose-Einstein condensate. 

\subsection{Contribution from external heating}
\label{subsec:heating_ME}
The external heating is described as a coupling of the center-of-mass mode to an effective bath with mean occupation number $\bar{N}$ \cite{Henkel_APB99,Brownnutt_RMP15}. The corresponding contribution to the master equation reads
\begin{equation}
    \begin{split}
        &\frac{\gamma}{2}(\bar{N}+1)\big(2\hat{a}_\mathrm{c.m.}\hat{\rho}_R\hat{a}_\mathrm{c.m.}^\dag-\hat{a}_\mathrm{c.m.}^\dag\hat{a}_\mathrm{c.m.}\hat{\rho}_R-\hat{\rho}_R\hat{a}_\mathrm{c.m.}^\dag\hat{a}_\mathrm{c.m.}\big)\\
        &+\frac{\gamma}{2}\bar{N}\big(2\hat{a}_\mathrm{c.m.}^\dag\hat{\rho}_R\hat{a}_\mathrm{c.m.}-\hat{a}_\mathrm{c.m.}\hat{a}_\mathrm{c.m.}^\dag\hat{\rho}_R-\hat{\rho}_R\hat{a}_\mathrm{c.m.}\hat{a}_\mathrm{c.m.}^\dag\big),
    \end{split}
    \label{eq:heating}
\end{equation}
where we omitted the time dependencies. For a bath at room temperature, we have $\bar{N}\gg1$, and the prefactor $\gamma(\bar{N}+1)\approx\gamma\bar{N}$ can be obtained empirically from experiments.

\section{Results}
\label{sec:results}
In this section, we report the results obtained by solving the master equation described in Eq.~\eqref{eq:equation_BEC}. The latter is used to study the dynamics of the ion chain and fidelity of the quantum gate in the presence of external heating and the ultracold gas. We refer to Appendix~\ref{app:gate} for some details about the behavior of the quantum gate and the definition of process fidelity. 

Unless stated differently, we fix the atom-ion $s$-wave scattering length at $a_\mathrm{ai}\simeq R^\star$. The ions are $^{174}$Yb$^+$, while the bath is an ultracold gas of either $^6$Li (Fermi) or $^7$Li (Bose) with density $n=10^{13}\,\mathrm{c.m.}^{-3}$ and temperature $T=200\,\mathrm{nK}$. For the ion chain, we consider the frequencies in Tab.~\ref{tab:frequencies}. Note that this choice would allow us to neglect the stretching (st) and wobbling (wb) mode in the description of the gate operation, strongly reducing the computational effort. In fact, $\delta_\mu\gg\Omega_\mu$ for said modes (see Sec.~\ref{subsec:ion_chain}). The gate, therefore, relies on the entanglement of the spin with the center-of-mass mode and the gate time is given by $2\pi/\delta_\mathrm{c.m.}=0.25\,\mathrm{ms}$. However, the wobbling mode has to be taken into account in order to properly describe the motion of the central ion inside the bath. The sums over the modes $\mu$ in all of the derived equations will hence be restricted to $\mu=\mathrm{c.m.},\mathrm{wb}$.

Moreover, we remark that the temperature of the ion chain is defined as
\begin{equation}
    T_\mathrm{chain}=\frac{1}{k_B}\frac{1}{2M}\sum_{\mu}\langle\hat{p}_\mu^2\rangle,
\end{equation}
where $\hat{p}_\mu=i\sqrt{\hbar M\omega_\mu/2}(\hat{a}_\mu^\dag-\hat{a}_\mu)$ is the momentum operator corresponding to mode $\mu$. Note that, in neglecting the stretching mode, we are not considering its contribution to the chain temperature. However, this is uncoupled from the other modes and is not affected by either heating or the quantum gas, meaning that it would add to the total temperature as a mere shift.
\begin{table}
    \centering
    \setlength{\extrarowheight}{1ex}
    \setlength{\tabcolsep}{5ex}
    \begin{tabular}{l l}
    \hline\hline
        Normal mode & Parameters \\[1ex]
    \hline
        \multirow[t]{3}{*}{c.m.:} & $\omega_\mathrm{c.m.}=2\pi\cdot500\,\mathrm{kHz}$\\
        & $\delta_\mathrm{c.m.}=2\pi\cdot4\,\mathrm{kHz}$\\
        & $\Omega_\mathrm{c.m.}=2\pi\cdot2\sqrt{3}\,\mathrm{kHz}$\\[1ex]
        \hline
        \multirow[t]{3}{*}{wobbling:} & $\omega_\mathrm{wb}=\omega_\mathrm{c.m.}\sqrt{29/5}$\\
        & $\delta_\mathrm{wb}=\omega_R-\omega_\mathrm{wb}$\\
        & $\omega_\mathrm{wb}=\Omega_\mathrm{c.m.}\sqrt{\omega_\mathrm{c.m.}/\omega_\mathrm{wb}}$\\[1ex]
        \hline\hline
    \end{tabular}
    \caption{Frequencies of center-of-mass and wobbling mode. The Raman beatnote frequency is $\omega_R=\delta_\mathrm{c.m.}+\omega_\mathrm{c.m.}$, by definition of $\delta_\mathrm{c.m.}$.}
    \label{tab:frequencies}
\end{table}
\subsection{Cooling dynamics}
\label{subsec:results_T}
\begin{figure}
    \centering
    \includegraphics[width=.45\textwidth]{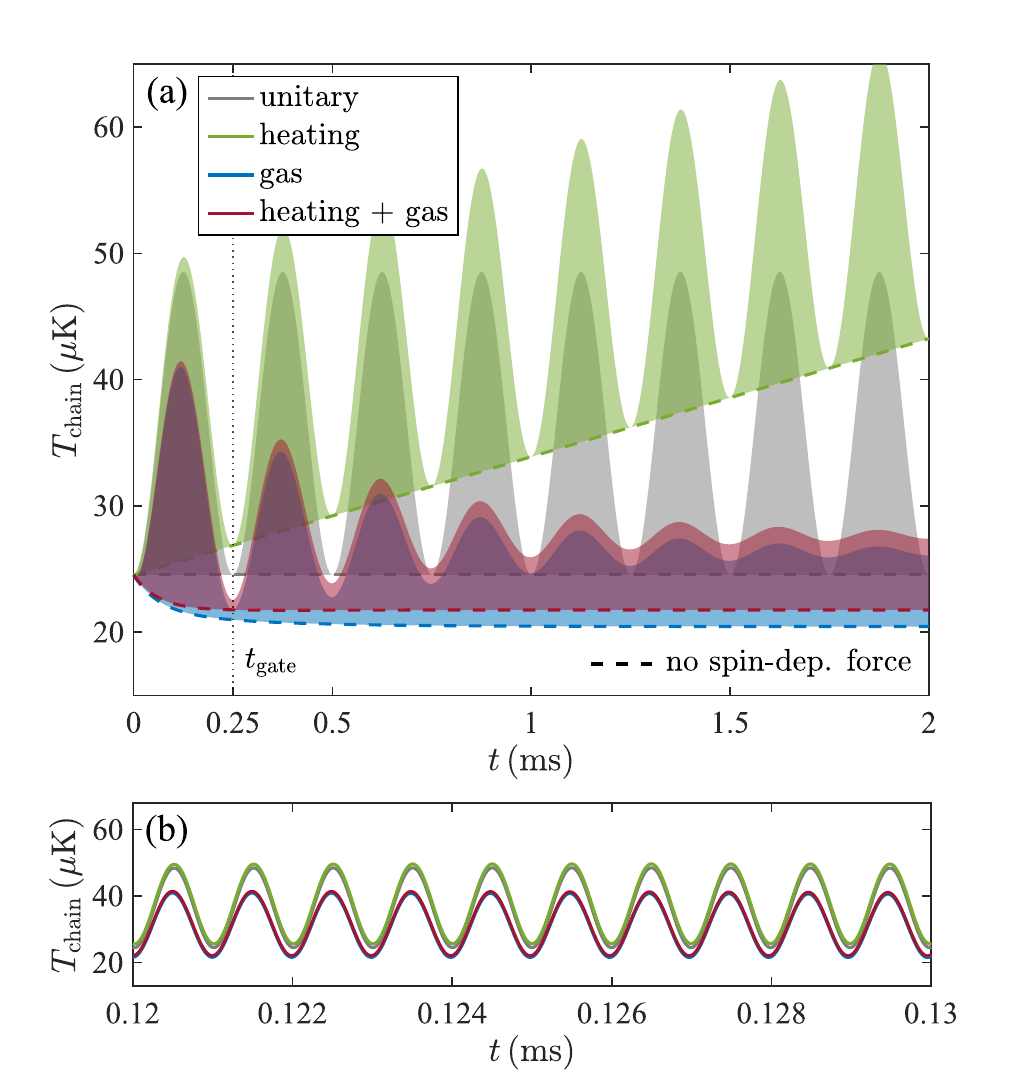}
    \caption{Temperature of ion chain as a function of time. The outer ions are initialized in the state $\ket{\uparrow\uparrow}$, while the central ion is initialized in an excited motional state (see text). (a) Comparison between unitary evolution (gray), in the presence of external heating with $\gamma\Bar{N}=200\,\mathrm{s}^{-1}$ (green), with coupling to the quantum gas (blue) and with all the contributions included (red). Dashed lines indicate the evolution without spin-dependent force. The vertical dotted line indicates the time $t_\mathrm{gate}=0.25\,\mathrm{ms}$. (b) Zoom on the region between $t=0.12$ and $0.13\,\mathrm{ms}$ showing the fast oscillation with frequency $\omega_\mathrm{c.m.}+\delta_\mathrm{c.m.}/2$.}
    \label{fig:T_ion}
\end{figure}
We consider the temperature of the ion chain and we study how this is affected by the coupling to the ultracold quantum gas and to the external heating. The spin of the two outer ions is set to the state $\ket{\uparrow\uparrow}$, while the central ion is initialized in the approximate thermal motional state $c_0\ket{0}\bra{0}_\mu+c_1\ket{1}\bra{1}_\mu$ for both the center-of-mass and wobbling mode, with $c_0=0.9$ and $c_1=0.1$. Here, $\ket{n}_\mu$ represents the $n$-th oscillator state of mode $\mu$, which contributes to the initial temperature with $c_n\hbar\omega_\mu(n+1/2)/(2k_\mathrm{B})$, where the factor $1/2$ is due to the equipartition theorem. 

In Fig.~\ref{fig:T_ion}(a), we compare the behavior of the temperature in different scenarios. We first note that, in absence of spin-dependent force acting on the outer ions, the time evolution of $T_\mathrm{chain}$ is monotonous (dashed lines). 
On the other hand, when the spin-dependent force explained in Sec.~\ref{subsec:ion_chain} is considered, each mode is driven with frequency $\omega_R$ and the ion temperature oscillates accordingly. In particular, we observe that this oscillation has a slow and a fast component [see Fig.~\ref{fig:T_ion}(b)], corresponding to the momentum of the center-of-mass mode oscillating with frequencies $\delta_\mathrm{c.m.}/2$ and $\omega_\mathrm{R}+\delta_\mathrm{c.m.}/2$, respectively~\cite{Kim:2011}. We remark that the wobbling mode has an analogous behavior, but its contribution to the oscillation of $T_\mathrm{chain}$ is not perceptible due to the much weaker driving strength compared to the center-of-mass mode.
In the case of unitary evolution (gray), the system occupies the same state after every gate cycles. Hence, the temperature retrieves its initial value at every multiple of $t_\mathrm{gate}$. In the presence of heating (green) a similar behavior is observed. In this case, however, the temperature increases linearly with a rate proportional to $\gamma\Bar{N}$. 
Interestingly, when the chain is coupled to the ultracold quantum gas, the latter guarantees the cooling of the chain. The dashed blue line shows that the ion temperature without spin-dependent force and without heating converges to a final value which is given by the ground state energy of the ion in the trap. In our case, it corresponds to $\hbar(\omega_\mathrm{c.m.}+\omega_\mathrm{wb})/(4k_B)\simeq20.45\,\mu\mathrm{K}$. 
Note that, if micromotion was considered, the ion would converge to a higher temperature \cite{Oghittu_PRA21}. 
When both the heating and ultracold gas are considered (red), a dynamical equilibrium is reached where the ion temperature stabilizes at a value which is higher than the ground state temperature. We observe that even in the case of high heating rate compared to typical values in experiments, the cooling action of the gas is remarkably fast. For the parameters considered in Fig.~\ref{fig:T_ion} and without spin-dependent force, the final temperature is reached in around $0.5\,\mathrm{ms}$, i.e. the time of two gate cycles. The blue and red shaded areas in Fig.~\ref{fig:T_ion}(a) show that the coupling to the gas damps the oscillation due to the spin-dependent driving force. In other words, the system behaves effectively as a driven damped harmonic oscillator and converges to a steady state where the temperature oscillates with frequency $\omega_\mathrm{R}$ and constant amplitude.

\subsection{Effect on process fidelity}
\label{subsec:results_F}
\begin{figure}
    \centering
    \includegraphics[width=.45\textwidth]{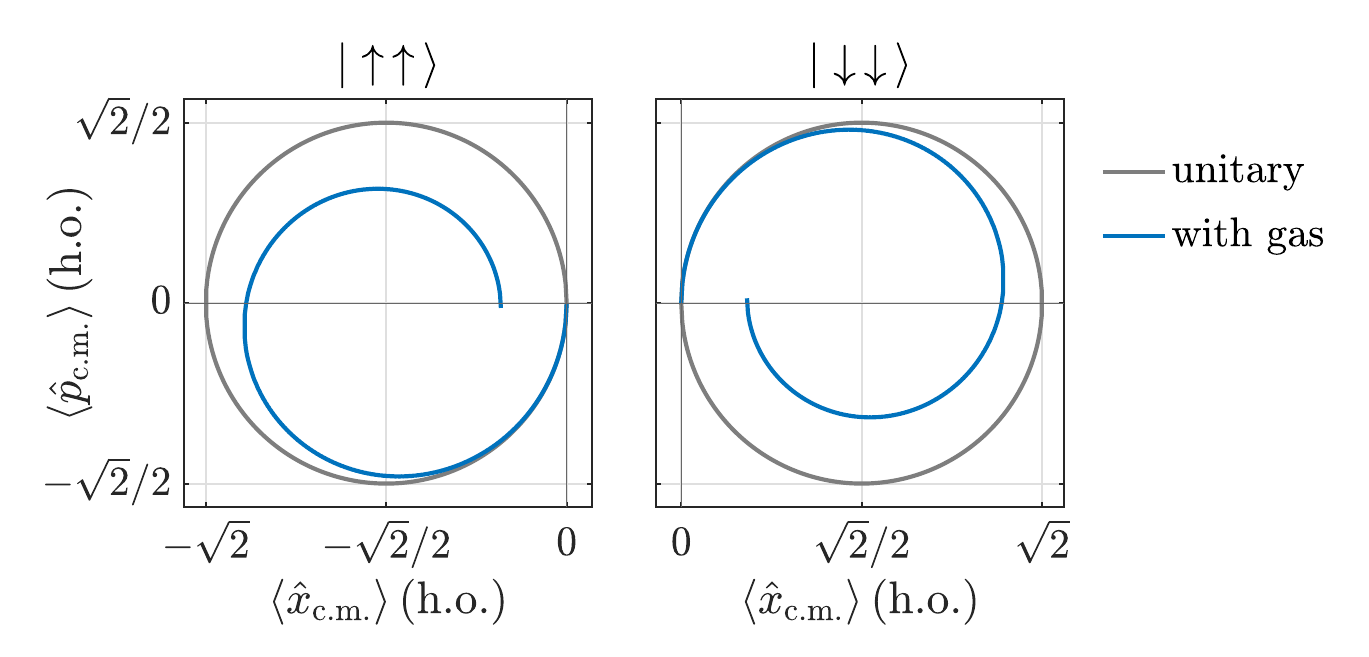}
    \caption{Phase space behavior of the center of mass mode in the frame rotating with $\omega_R$ for two different initial spin states. Gray lines correspond to unitary evolution; blue lines show motional decoherence of the system when coupled to the ultracold bath.}
    \label{fig:phase_space}
\end{figure}
The error induced on the gate operation by the coupling to the ultracold gas can be understood by observing the phase space behavior of the system. As shown in Fig.~\ref{fig:phase_space}, the unitary evolution (gray) in the frame rotating with $\omega_\mathrm{R}$ corresponds to a circle in the phase space of the center-of-mass mode. On the other hand, the presence of the gas (blue) deforms the trajectory and does not allow the circle to close after the gate time $t_\mathrm{gate}=0.25\,\mathrm{ms}$, indicating decoherence of motional states.
\begin{figure}
    \centering
    \includegraphics[width=.45\textwidth]{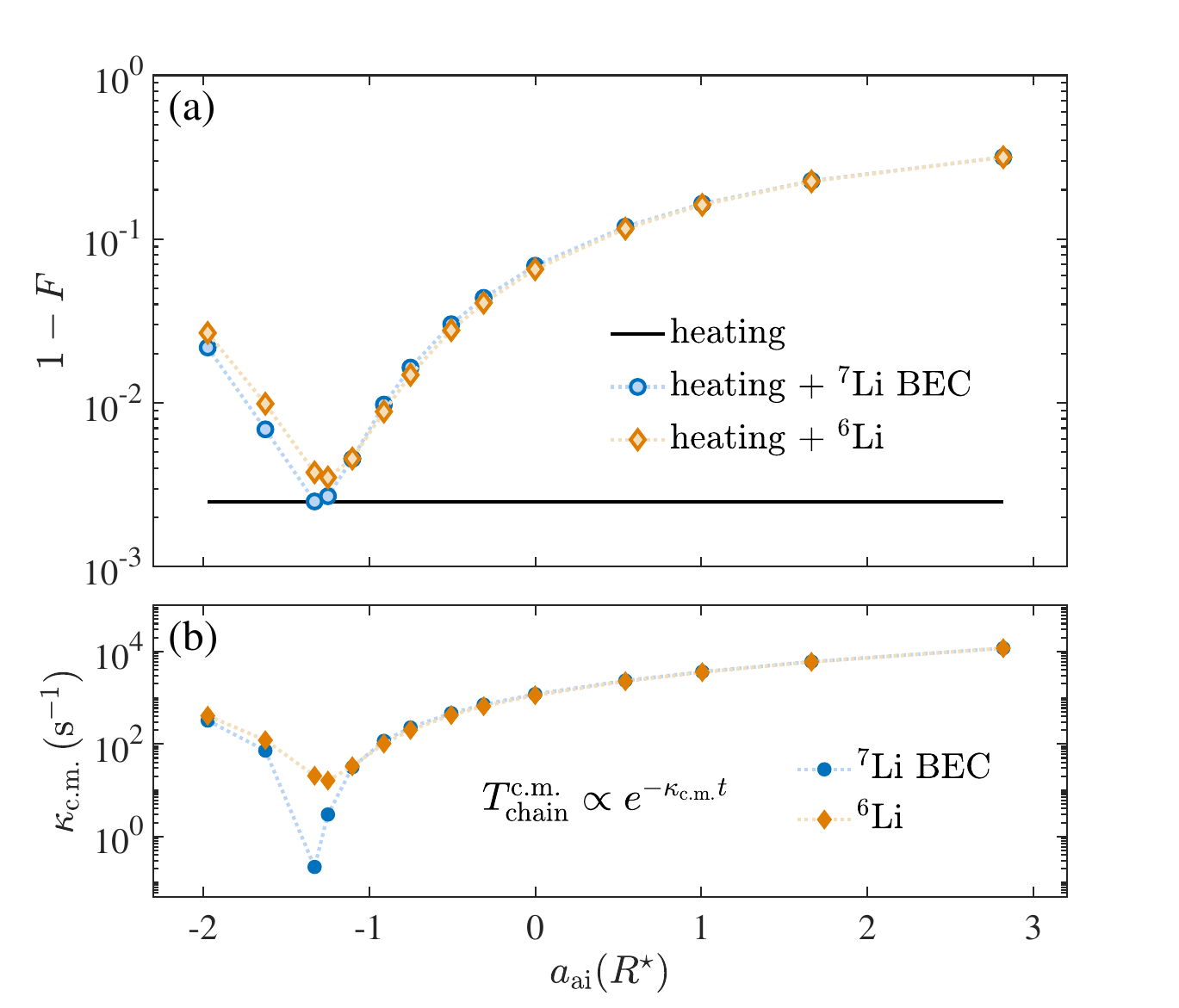}
    \caption{(a) Infidelity $1-F$ after single gate cycle with heating rate $\gamma\Bar{N}=20\,\mathrm{s}^{-1}$ and different values of the atom-ion scattering length. The black solid line indicates the value obtained without coupling to the ultracold gas. (b) Cooling rate of the center-of-mass mode as a function of scattering length. The values of $\kappa_\mathrm{c.m.}$ are obtained by fitting the temperature associated to the center-of-mass mode with an exponential function $Ae^{-\kappa_\mathrm{c.m.}t}+B$. All the dotted lines are mere guides to the eye.}
    \label{fig:F_ion}
\end{figure}
Nevertheless, this behavior can be mitigated by properly tuning the atom-ion scattering length. In Fig.~\ref{fig:F_ion}(a), we observe that for $a_\mathrm{ai}\simeq-1.3\,R^\star$, the effect of the gas on the process fidelity is strongly suppressed. The same behavior is observed for both a $^7$Li Bose-Einstein condensate (blue circles) as well as a spin-Polarized ultracold Fermi gas of $^6$Li atoms (orange triangles), although a better fidelity can be achieved in the bosonic environment. This difference may be due to the stronger coupling of the ion to a fermionic bath, which is also responsible for the difference in the cooling rates shown in Fig.~\ref{fig:F_ion}(b). In this regard, we note that the behavior of the infidelity gives indication of how much the center-of-mass motion couples to the atomic phonons in the gas. In Fig.~\ref{fig:F_ion}(b), we observe that the optimal regime for cooling corresponds to the values of $a_\mathrm{ai}$ for which the infidelity is higher. Finally, we note that the interaction tune-out value of the scattering length does not correspond to $a_\mathrm{ai}=0$ due to the interplay between molecular and confinement-induced resonances~\cite{Oghittu_PRA21}.

This result, combined with the cooling dynamics presented in Sec.~\ref{subsec:results_T}, may have interesting applications in experiments. There, the atom-ion scattering rate can be controlled by tuning a magnetic field close to a Feshbach resonance~\cite{Weckesser2021oof}, making it possible to alternate from the regime were the coupling to the bath is suppressed and gate operations are performed, to the one where the effect of the ultracold gas is maximized and the ions are cooled. On this regard, we finally note that three body recombination could affect the experimental realization of our model. This would reduce the lifetime of the central ion in the trap during the cooling phase due to the enhanced ion-bath coupling. Although the estimation of three body recombination time is beyond the scope of this paper, we can observe that it scales with the density of the gas as $n^{-2}$~\cite{Perez-Rios_MP21}, whereas the cooling time scales as $n^{-1}$. Therefore, the potential problem of ion loss can be mitigated efficiently by reducing the gas density.

\section{Conclusions and outlook}
\label{sec:conclusions}
We studied phonon-mediated interactions between ionic qubits coupled to an ultracold quantum gas and external heating. We derived a master equation for the reduced density matrix of the chain. We find that the presence of the atomic bath has a negative effect on the quality of quantum gates between the ion qubits. On the other hand, we calculate that the atomic bath can be used to keep the ion crystal cool in the presence of external heating due to electric field noise. Tuning of the atom-ion scattering length using a magnetic field allows one to set the cooling rate of the atoms. In particular, the adverse effects during quantum gates may be limited, while the buffer gas cooling may be maximized in between gates. This shows that buffer gas cooling is a viable technique for trapped ion quantum systems. We remark that due to the long range of the interaction between charged particles, also chains with more than three ions can be cooled by immersing a single one of them in the ultracold bath.

The system offers further opportunities in studying the decoherence of non-classical states of ion motion and spin-motion entangled states in the presence of a fermionic or bosonic quantum bath. The system may be used to explore the crossover between classical and quantum dynamics~\cite{Zurek:2003} and the occurrence of classical and quantum chaos in mixtures of ultracold atoms and trapped ions~\cite{Pinkas:2024}. Moreover, the coupling of the trapped ion atom mixture to qubits may allow for quantum-enhanced measurements of the properties of this system in analogy to e.g. Ref.~\cite{Hempel:2013}. There, a fragile spin-motion entangled state was used to detect the scattering of a single photon. The momentum transfer between an ultracold Li atom and an Yb$^+$ is comparable to that of scattering infrared photons such that the scheme could be extended to quantum enhanced detection of atom-ion scattering.


\section*{Acknowledgements}
This work was supported by the Dutch Research Council (Grant Nos. 680.91.120 and VI.C.202.051), (R.G.). A.S.N is supported by the Dutch Research Council (NWO/OCW), as part of the Quantum Software Consortium programme (Grant number 024.003.037), Quantum Delta NL (Grant number NGF.1582.22.030) and ENW-XL (Grant number OCENW.XL21.XL21.122). L.O. acknowledges the European Cooperation in Science \& Technology (COST Action Grant No. CA17113).


\appendix

\section{Time dependence of ion position}
\label{app:long_equations}
Here we provide the explicit form of the operator defined in Eq.~\eqref{eq:x_t-tau}, describing the motion of the central ion of the chain in the frame rotating with $\omega_R$ and in absence of the bath:
\begin{widetext}
\begin{equation}
    \begin{split}
        \hat{x}_R(t,-\tau)=\sum_{\mu}b_{2,\mu}\lambda_\mu&\bigg\{\hat{a}_\mu e^{-i\delta_\mu\tau}e^{-i\omega_Rt}+\hat{a}_\mu^\dag e^{i\delta_\mu\tau}e^{i\omega_Rt}\\
        +&\frac{\Omega_\mu}{2\delta_\mu}\sum_{j=1,3}b_{j,\mu}\hat{\sigma}_j\Big[e^{-i\delta_\mu\tau}e^{-i\omega_R t}+e^{i\delta_\mu\tau}e^{i\omega_R t}-2\mathrm{cos}\big(\omega_R t\big)\Big]\bigg\}.
    \end{split}
    \label{eq:x_t-tau_long}
\end{equation}
\end{widetext}
We remark that this result relies on the Hamiltonian $\hat{H}_\mathrm{chain}$ being independent of time, hence on the rotating wave approximation (Sec.~\ref{subsec:ion_chain}). In case of a time-dependent chain Hamiltonian, the calculation would require a different procedure, as in the case of the Paul-trapped ion~\cite{Leibfried:1996,Krych_PRA15}.

\section{Quantum Gate}
\label{app:gate}
The spin-dependent Hamiltonian described in Sec.~\ref{subsec:ion_chain} can be used to perform a geometric phase gate. In this section, we briefly summarize some of the basic features of this operation and we define the process fidelity.

\subsection{Gate behavior}
\label{subapp:gate_behavior}
Given an initial $4\times4$ matrix $\hat{\rho}_0$ describing the state of the two qubits, the final state after the gate operation in the case of an ideal process represented as $\Lambda$ is given in the $\{\ket{\uparrow}$, $\ket{\downarrow}\}$ basis by
\begin{equation}
    \Lambda(\hat{\rho}_0)=\hat{\mathcal{U}}^\dag(t_\mathrm{gate})\,\hat{\rho}_0\,\hat{\mathcal{U}}(t_\mathrm{gate}),
\label{eq:final_rho_ideal}
\end{equation}
with $\hat{\mathcal{U}}(t_\mathrm{gate})=\mathrm{Diag}\{1,-i,-i,1\}$. For instance, let us consider the initial state to be $\ket{++}\equiv\ket{+}_1\otimes\ket{+}_3$. The corresponding density matrix $\hat{\rho}_0=\ket{++}\bra{++}$ in the basis $\{\ket{\uparrow}$, $\ket{\downarrow}\}$ is the $4\times4$ matrix where all the elements are equal to $1/4$. According to Eq.~\eqref{eq:final_rho_ideal}, the final density matrix is
\begin{equation}
    \Lambda(\hat{\rho}_0)=\begin{pmatrix}
        \frac{1}{4} & \frac{i}{4} & \frac{i}{4} & \frac{1}{4}\\[1ex]
        -\frac{i}{4} & \frac{1}{4} & \frac{1}{4} & -\frac{i}{4}\\[1ex]
        -\frac{i}{4} & \frac{1}{4} & \frac{1}{4} & -\frac{i}{4}\\[1ex]
        \frac{1}{4} & \frac{i}{4} & \frac{i}{4} & \frac{1}{4}
    \end{pmatrix},
\label{eq:final_matrix_example}
\end{equation}
which corresponds to the state $(\ket{++}-i\ket{--})/\sqrt{2}$.

\subsection{Process fidelity}
The operation explained above refers to an ideal gate. However, the coupling to the external heating and to the quantum gas introduces an error. To quantify how close the non-ideal process, indicated as $\Gamma$, reproduces the ideal one, we define the process fidelity as \cite{Greenway_PRR21}
\begin{equation}
    F(\Lambda,\Gamma)=\frac{1}{d^2}\sum_{i=1}^{d^2}\mathrm{Tr}\Big\{\Lambda(\hat{\rho
    }_i^\dag)\Gamma(\hat{\rho}_j)\Big\},
\end{equation}
where $\{\hat{\rho}_i\}$ is a complete set of mutually orthonormal operators in the Hilbert space of the two qubits with dimension $d=4$. In the case of a two-qubit gate, a natural choice is to define each operator $\hat{\rho}_i$ as one of the possible tensor products $\hat{\sigma}_1^\xi\otimes\hat{\sigma}_3^\zeta$, where $\hat{\sigma}_j^\xi$ ($\xi=0,x,y,z$) are the Pauli matrices in the subspace of ion $j$. Note that these operators do not represent, in general, a quantum state. Hence, the experimental assessment of $F$ requires some additional precautions. We refer the interested reader to Ref.~\cite{Greenway_PRR21} for a complete treatment of the topic.

\bibliography{biblio_RG}

\end{document}